\documentclass[
  aps,
  pre,
  preprint,
  superscriptaddress,
  floatfix
]{revtex4-2}

\usepackage{amsmath,amssymb,amsfonts}
\usepackage{graphicx}
\usepackage{booktabs}
\usepackage{comment,color}

\begin{document}

\title{A Reexamination of Noise-Driven Robustness Evolution in Gene Regulatory Networks}

\author{Momoko Amemiya}
\affiliation{Graduate School of Frontier Sciences,
The University of Tokyo}
\affiliation{Department of Information Science, Ochanomizu University}

\author{Ayaka Sakata}
\affiliation{Department of Information Science, Ochanomizu University}
\affiliation{RIKEN Center for AIP}

\date{\today}

\begin{abstract}

Gene regulatory network (GRN) models are widely used to investigate gene-expression dynamics in biological systems.
In this study, we re-examine an evolutionary GRN model proposed by Kaneko, which was used to investigate the relationship between robustness to nongenetic phenotypic noise and robustness to mutation. 
We explicitly reconstruct the simulation protocol by specifying computational details omitted from the original study and reproduce its main qualitative findings, 
including the noise-dependent suppression of low-fitness individuals, the relationship between isogenic phenotypic variance and genetic variance, and the increased robustness of evolved networks under certain expression noise.
We further analyze the robustness of evolved populations to variations in initial conditions. While the representative-network analysis shows quantitative differences from the original study, the population-level analysis supports the qualitative conclusion that evolution under higher phenotypic noise leads to broader basins of attraction.
We provide a reusable implementation and publicly release the source code to support reproducibility and facilitate further studies on robustness in dynamical GRN models. 

\end{abstract}

\maketitle

\section{Introduction}

Living systems consist of many interacting components, including genes, proteins, cells, and organisms.
Through nonlinear interactions among these components, biological systems exhibit collective behaviors such as adaptation, self-organization, and information processing.
Because of these properties, living systems have often been studied as complex systems not only in biology, but also in physics, engineering, and information science.
Since the development of systems biology, such perspectives have been further extended to the analysis of large-scale biological networks \cite{kitano2002systems}.
Moreover, biological systems have been analyzed as information-processing systems in which network structures and dynamical patterns play essential roles in cellular functions and decision making \cite{alon2019introduction}.
These perspectives have motivated interdisciplinary studies aimed at identifying universal principles underlying living systems.

Among the various universal properties discussed in biological complex systems, robustness has attracted particular attention.
Biological organisms are able to maintain stable functions and developmental processes being exposed to environmental fluctuations and internal noise.
For example, cells can preserve specific gene expression patterns despite stochastic fluctuations in molecular interactions \cite{kaern2005stochasticity}. 
Similarly, multicellular organisms can achieve reliable body plans during embryonic development even under genetic and environmental perturbations \cite{waddington2014strategy}. 
At a larger scale, ecosystems are also known to maintain their functions and biodiversity against external disturbances through complex interactions among species \cite{holling1973resilience}. 
These examples suggest that robustness is a universal property observed across multiple levels of biological organization, rather than being limited to a specific biological process.
Understanding how such stability emerges from complex interactions is an important problem in modern science.

Biological systems often involve highly complex interactions that are difficult to directly observe and analyze, theoretical models provide useful frameworks for complementing experimental data and for extracting phenomenological and universal principles underlying biological phenomena.
Over the past decades, various approaches based on dynamical systems theory, statistical physics, and network science have been proposed to investigate how robustness, stability, and adaptability emerge in living systems \cite{kitano2004biological, bornholdt2008boolean}.
Among these approaches, gene regulatory network (GRN) models have been widely used to study the dynamical behavior of gene expression and cellular states.
In GRN models, genes and their regulatory interactions are represented as network structures, and cellular dynamics emerge from collective interactions among many components \cite{alon2019introduction,kauffman1992origins}.

\subsection{Kaneko's dynamical GRN model}

Among the various forms of biological robustness, those against dynamical fluctuations and mutations have attracted particular attention.
Biological systems are required not only to maintain stable functions under stochastic noise and environmental perturbations, but also to preserve their functionalities against genetic changes accumulated through evolution. 
The relationship between different forms of robustness has therefore been extensively discussed in evolutionary and systems biology \cite{wagner2005robustness}.

To investigate the relationship between different forms of robustness, Kaneko proposed an evolutionary gene regulatory network (GRN) model \cite{Kaneko2007}. 
In this model, robustness against phenotypic noise and robustness against mutations 
were investigated by combining the dynamics of gene expression with evolutionary processes.
The study suggested that these two types of robustness are closely related and can emerge through evolutionary processes under noisy gene-expression dynamics.

\subsection{Our contribution}

Kaneko's study provided a pioneering demonstration of the relationship between different forms of robustness in evolutionary GRN models.
Subsequent studies further investigated the relationships among phenotypic fluctuations, mutational robustness, and evolutionary dynamics using related models \cite{sakata2009funnel,kaneko2022evolution}.
However, reproducing the original simulations is not straightforward because the model involves many parameters and implementation details, some of which are not fully specified in the original paper.

More generally, in many simulation-based studies in physics for complex systems research from the 1980s to the early 2000s, it was not yet common practice to publicly release source code or detailed computational procedures. As a result, reconstructing the simulation conditions and reproducing published results solely from the descriptions in the literature can be difficult.

The purpose of this study is to reproduce the main results of Kaneko \cite{Kaneko2007} by explicitly specifying simulation procedures and parameter settings that were not fully described in the original study. 
In addition, we organize the implementation in a reusable form and publicly release the source code\cite{Amemiya_GRN_code}.

Our simulations reproduce the main qualitative behaviors reported in the original study, including the noise-dependent suppression of low-fitness individuals and the relationship between robustness to phenotypic noise and robustness to mutation. 
In addition, we analyze the stability of evolved populations against variations in initial conditions. 
Populations evolved under larger noise strengths tend to exhibit more stable final expression patterns and lower variability in fitness outcomes across different initial conditions. 
These findings suggest that noise-dependent changes in dynamical basin structures emerge not only in individual representative networks but also across the population.

Overall, this study provides a reproducible implementation of a foundational dynamical GRN model. By providing detailed computational procedures and parameter settings, we facilitate further investigations of robustness and evolutionary dynamics based on this model.

\section{Model}

Biologically, a gene regulatory network represents regulatory interactions among genes and their products, which control gene expression and cellular functions.
Such regulatory networks play important roles in cellular dynamics, development, and morphogenesis.
In mathematical and computational studies, GRNs are often represented as simplified dynamical systems in which genes and their regulatory interactions are modeled using networks and dynamical equations. 
Kaneko's model studied in this work is based on such an abstraction and describes gene expression dynamics through interactions among genes.

Furthermore, Kaneko's model considers a population of model cells, each  
characterized by its own GRN structure.
Through mutation and selection processes acting on the regulatory network, the population evolves over generations.
In this framework, the regulatory network structure corresponds to the genotype, whereas the resulting gene expression pattern corresponds to the phenotype. 
The phenotype is dynamically generated through intracellular dynamics and can fluctuate due to stochastic noise in gene expression.

\subsection{Mathematical expression}

\paragraph{Gene expression dynamics}

Here, we consider a GRN consisting of $M$ genes. 
The expression state of the $i$-th gene is denoted by $x_i\in\mathbb{R}$.
Positive expression states ($x_i>0$) are interpreted as the ``on" state of a gene, whereas zero or negative values ($x_i\leq0$) correspond to the ``off" state.
The regulatory interaction from gene $j$ to gene $i$ is represented by $J_{ij}\in\{-1,0,+1\}$. Positive values of $J_{ij}$ indicate that expression of gene $j$ tends to enhance the expression of gene $i$, whereas negative values indicate inhibitory effects. A value of $J_{ij}=0$ indicates that there is no direct regulatory interaction between the two genes.
In this model, the interaction matrix $J$, which determines the gene expression dynamics, represents the genotype of an individual. The resulting gene expression pattern generated by the dynamics represents the phenotype.

Among the $M$ genes, the first $K$ genes are defined as output genes that determine the fitness of the system. 
These output genes correspond to genes whose expression patterns directly affect fitness evaluation. The remaining genes contribute to the internal regulatory dynamics.

The time evolution of gene expression levels is not determined solely by the interaction matrix $J$, because gene expression processes intrinsically involve stochastic fluctuations arising from molecular interactions inside the cell. 
To incorporate 
stochastic phenotypic fluctuations, the dynamics are modeled by the following stochastic differential equation:
\begin{equation}
\frac{dx_i}{dt}
=
-x_i
+
\tanh\left(\beta \sum_{j>K}^{M} J_{ij} x_j \right) %
+
\sigma \eta_i(t),
\label{eq:grn_dynamics}
\end{equation}
where $\beta$ controls the steepness of the sigmoid response function, $\sigma$ represents the noise strength, and $\eta_i(t)$ denotes independent Gaussian white noise satisfying $\langle \eta_i(t)\eta_j(t') \rangle
=
\delta_{ij}\delta(t-t')$.
In Kaneko's model, 
regulatory inputs to each gene are provided only by non-output genes; that is, the summation in eq.\eqref{eq:grn_dynamics} excludes the output genes used for fitness evaluation.
Although $\{x_i\}$ are defined as continuous variables over $\mathbb{R}$, the sigmoid response function causes the dynamics to be effectively attracted to the range $[-1,+1]$, with small deviations arising from the noise term.

\paragraph{Definition of fitness}

The fitness of each individual is determined from the expression patterns of the $K$ output genes. 
In the present model, the target functional state is defined such that all output genes are in the ``on" state, namely $x_i>0$ for $i=1,\dots,K$.
This setting reflects the idea that biological fitness is often determined not by the states of all genes, but by a limited number of functionally important outputs, such as key regulatory or downstream functional genes \cite{levine2005gene}.

For simplicity, we assume this target expression pattern, although
other target expression patterns are expected to lead to qualitatively equivalent behaviors due to the symmetry of the model.
Under this setting, the fitness of genotype $J$ is evaluated based on the gene-expression dynamics generated by $J$.
More specifically, the fitness is defined from the time average of the output-gene states:
\begin{equation}
F(J)
=\sum_{j=1}^{K}\left(\langle S(x_j)\rangle_{\mathrm{temp}}-1\right).
\label{eq:fitness}
\end{equation}
Here,
\[
S(x)=
\begin{cases}
1 & (x>0),\\
0 & (x\le 0),
\end{cases}
\]
and \(\langle\cdot\rangle_{\mathrm{temp}}\) denotes the time average from \(T_{\mathrm{ini}}\) to \(T_f\):
\[
\langle S(x_j)\rangle_{\mathrm{temp}}=\frac{1}{T_f-T_{\mathrm{ini}}}\int_{T_{\mathrm{ini}}}^{T_f} S(x_j(t)) dt.
\]
Here, \(x_j(t)\) denotes the expression level of gene \(j\) at time \(t\), governed by \eqref{eq:grn_dynamics}, and \(T_{\mathrm{ini}}\) is introduced in order to exclude the initial transient dynamics.

Under this definition, fitness takes its maximum value \(F=0\) when all output genes remain ``on" throughout the measurement period.

\paragraph{Evolutionary process}

Using the fitness function defined above, the genotype $J$ evolves through an evolutionary process. 
Even for the same genotype, stochastic gene-expression dynamics can generate different phenotypic trajectories and consequently different fitness values.
Therefore, the fitness used for selection
is evaluated by averaging over multiple independent realizations of the dynamics.
Specifically, we define the mean fitness of each genotype as the average fitness over $L$ independent realizations of the stochastic gene-expression dynamics:
\begin{align}
\bar{F}(J)
=
\frac{1}{L}\sum_{\ell=1}^{L} F^{(\ell)}(J),
\label{eq:mean_fitness}
\end{align}
where \(F^{(\ell)}(J)\) denotes the fitness obtained in the \(\ell\)-th trial under the genotype $J$.

At each generation, individuals are selected according to their mean fitness $\overline{F}$, after which mutations are introduced into the interaction matrix \(J\) to generate the next-generation population.

\section{Quantities for robustness evaluation}

In the present model, variability arises from both genetic and phenotypic sources.
Genetic variability arises from differences among individuals in the evolving population, whereas phenotypic variability emerges from stochastic gene-expression dynamics under a given genotype.

Kaneko \cite{Kaneko2007} introduced two types of variance arising from two different sources of variability.
The first is the isogenic phenotypic variance \(V_{ip}\), which measures  
variation in fitness
caused by stochastic noise under a fixed genotype. 
The second is the genetic variance \(V_g\), which measures variation in mean fitness arising from differences among genotypes.

Below, we introduce the definitions of \(V_{ip}\) and \(V_g\).

\subsection{Isogenic phenotypic variance \(V_{ip}\)}

Using the fitness \(F^{(\ell)}(J)\) obtained from the \(\ell\)-th realization of the stochastic expression dynamics under genotype \(J\), the isogenic phenotypic variance is defined as
\[
V_{ip}(J)
=
\frac{1}{L}
\sum_{\ell=1}^{L}
\left(
F^{(\ell)}(J)
-
\overline{F}(J)
\right)^2,
\]
where $\overline{F}(J)$ is the mean fitness given by \eqref{eq:mean_fitness}.
Smaller values of \(V_{ip}(J)\) indicate that fitness is less influenced by stochastic fluctuation under a fixed genotype.
Since these fitness variations reflect the phenotypic fluctuations induced by stochastic gene-expression dynamics,
\(V_{ip}\) serves as a measure of robustness against such noise.

The population-averaged isogenic phenotypic variance is then defined as
\[
V_{ip}
=
\langle V_{ip}(J) \rangle_J,
\]
where \(\langle \cdot \rangle_J\) denotes averaging over genotypes in the population.

\subsection{Genetic variance \(V_g\)}

Different genotypes generate different expression dynamics and consequently different fitness values. 
The genetic variance \(V_g\) measures fitness variation originating from genotype differences.
Using the genotype-dependent mean fitness \(\overline{F}(J)\), the genetic variance is defined as
\[
V_g
=
\left\langle
\left(
\overline{F}(J)
-
\langle \overline{F}(J) \rangle_J
\right)^2
\right\rangle_J.
\]
We note that $\langle \overline{F}(J) \rangle_J$ denotes the population average of the mean fitness.
In the following, we refer to this quantity simply as the population-averaged fitness.
Smaller \(V_g\) indicate that genetic variation produces smaller changes in fitness. 
Since these fitness variations result from genetic differences, $V_g$ serves as a measure of robustness against mutation.

\subsection{Relationship between \(V_{ip}\) and \(V_g\)}

Kaneko \cite{Kaneko2007} showed that the genetic variance tends to remain smaller than the isogenic phenotypic variance under sufficiently large gene-expression noise:
\[
V_g < V_{ip}.
\]
This relationship suggests that robustness against mutation and robustness against stochastic noise, although arising from distinct sources of phenotypic variability, are closely linked through evolution. 
The study further showed that this relationship emerges only when the noise strength in the gene-expression dynamics exceeds a certain threshold.

\section{Setting of numerical simulation}
\begin{table}[ht]
\caption{Main simulation parameters used in this study.}
\label{tab:param}
\begin{ruledtabular}
\begin{tabular}{ll}
Parameter & Value \\
\hline
Number of genes \(M\) & 64 \\
Number of output genes \(K\) & 8 \\
\(\beta\) & 7 \\
Noise strength \(\sigma\) & 0.005--1.00 \\
Trials per individual \(L\) & 300 \\
Time step \(dt\)~* & 0.005 \\
Relaxation time \(T_{\mathrm{ini}}\)~* & 80 \\
Population size \(N\) & 300 \\
Measurement period \(T_f - T_{\mathrm{ini}}\)~* & 10 \\
Elite fraction \(r_{\mathrm{elite}}\) & 0.25 \\
Number of mutations per generation $n_m$~* & 1 \\
Initial interaction probability $p_0$~* & 0.5  \\
\end{tabular}
\end{ruledtabular}
\end{table}

To reproduce and examine Kaneko's findings 
on the role of gene-expression noise in the evolution of robustness,
we perform simulations for various values of the noise strength \(\sigma\).
Table~\ref{tab:param} summarizes the main parameters used in the model. 
Parameters marked with * were not specified in the original study \cite{Kaneko2007} and were chosen in this work to  enable reproducible simulations. 
We chose $T_f-T_{ini}=10$ for computational efficiency, having verified that longer measurement periods did not change the qualitative behavior of the model.

Simulations are conducted for a range of noise strengths
$[0.005,1]$, 
under the same numerical settings described in the following subsections.
For comparison across noise strengths, the same random seed is used for all simulations.

\subsection{Gene expression dynamics}

For each genotype $J$, we perform $L$ independent trials of the stochastic expression dynamics described by Eq.~\eqref{eq:grn_dynamics}.
Each simulation trial is initialized with the same expression state,
\[
x_i(0)=-1
\qquad (i=1,\dots,M),
\]
corresponding to all genes initially being in the ``off" state.
Because Eq.~\eqref{eq:grn_dynamics} is a stochastic differential equation,
numerical integration is performed using the Euler--Maruyama method.
After discarding an initial transient period,
the resulting expression trajectories are
used to evaluate fitness.

\subsection{Evolutionary dynamics}

We perform evolutionary simulations on a population consisting of \(N\) genotypes. 
The initial population consists of $N$ randomly generated genotypes where each interaction $J_{ij}$ is independently set to a nonzero value with probability $p_0$. 
To ensure comparability across noise conditions, the same initial population is used for all values of the noise strength \(\sigma\).
At each generation, the evolutionary process consists of three steps: evaluation of fitness, selection, and mutation.
First, the mean fitness \(\overline{F}(J)\) of each genotype in the population is evaluated based on the stochastic expression dynamics \eqref{eq:grn_dynamics}.
Second, genotypes are ranked according to their mean fitness, and the top fraction \(r_{elite}\) of the population is retained as parents. The parameter \(r_{elite}\) controls the selection pressure.
Third, \(\frac{1}{r_{elite}}\) offspring are generated from each selected genotype through mutation, so that the population size is maintained across generations. 
Mutation is implemented by selecting \(n_m\) elements \(J_{ij}\) and replacing each selected element by one of the other two possible values.

\section{Reproduction of Previous Results}

Here, we summarize the main results reported in \cite{Kaneko2007} that we aim to reproduce.

\begin{description}
    \item[\textbf{Result 1 (Noise strength and the top fitness)}] The maximum fitness in the population increases more rapidly under low-noise conditions during early generations.
    \item[\textbf{Result 2 (Noise strength and the worst fitness)}] When the noise strength is sufficiently large, the minimum fitness in the population increases over generations. In contrast, under low-noise conditions, the minimum fitness remains low throughout generations. 
    \item[\textbf{Result 3 (Fitness distributions depend on noise strengths)}]
    At sufficiently large noise strengths, the population mean fitness averaged over generations approaches the maximal value, and the worst fitness also approaches the maximal value.
    In contrast, under low-noise conditions, the worst fitness remains substantially lower, and low-fitness individuals persist in the population, resulting in a reduced population mean fitness. Consistent with these observations, the fitness distribution in the final generation is broadly spread toward low-fitness values under low-noise conditions, whereas it becomes sharply concentrated near the maximal fitness under sufficiently large noise.
    \item[\textbf{Result 4 (Noise threshold)}]
    A threshold noise strength was observed that separates a regime in which low-fitness individuals persist from a regime in which nearly the entire population evolves toward the maximum-fitness state.
    \item[\textbf{Result 5 (Relationship between $V_g$ and $V_{ip}$)}]
    Above the threshold noise strength, the inequality \(V_g <V_{ip}\) is maintained throughout evolution. As the noise strength is reduced toward the threshold, \(V_g\) increases and approaches \(V_{ip}\), eventually becoming comparable in magnitude. Furthermore, after sufficient evolutionary time above the threshold noise strength, \(V_g\) becomes approximately proportional to \(V_{ip}\) while maintaining the relationship \(V_g<V_{ip}\).
    \item[\textbf{Result 6 (Dynamical robustness} and basin expansion)]
    Networks evolved under 
    large noise strengths exhibited greater robustness of gene-expression dynamics to stochastic fluctuations, maintaining the target expression state even under strong noise. They also reached the target state from a larger fraction of random initial conditions, indicating an expansion of the basin of attraction associated with the target state.

\end{description}

\subsection{Fitness distribution depends on noise strength}

\begin{figure}[ht]
  \centering
  \begin{minipage}{0.48\linewidth}
    \centering
    \includegraphics[width=\linewidth]{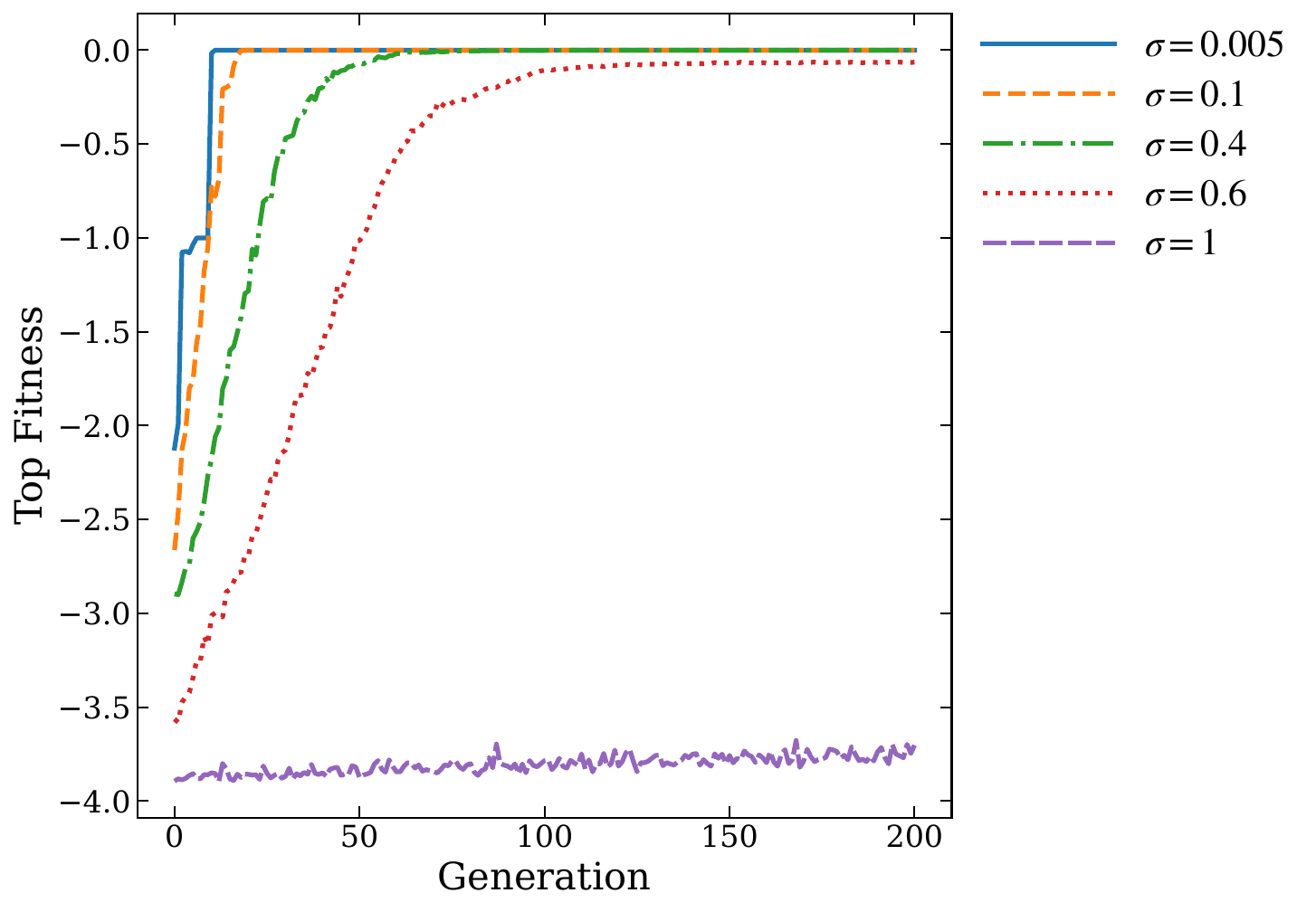}
    \small (a) Top Fitness
  \end{minipage}
  \hfill
  \begin{minipage}{0.48\linewidth}
    \centering
    \includegraphics[width=\linewidth]{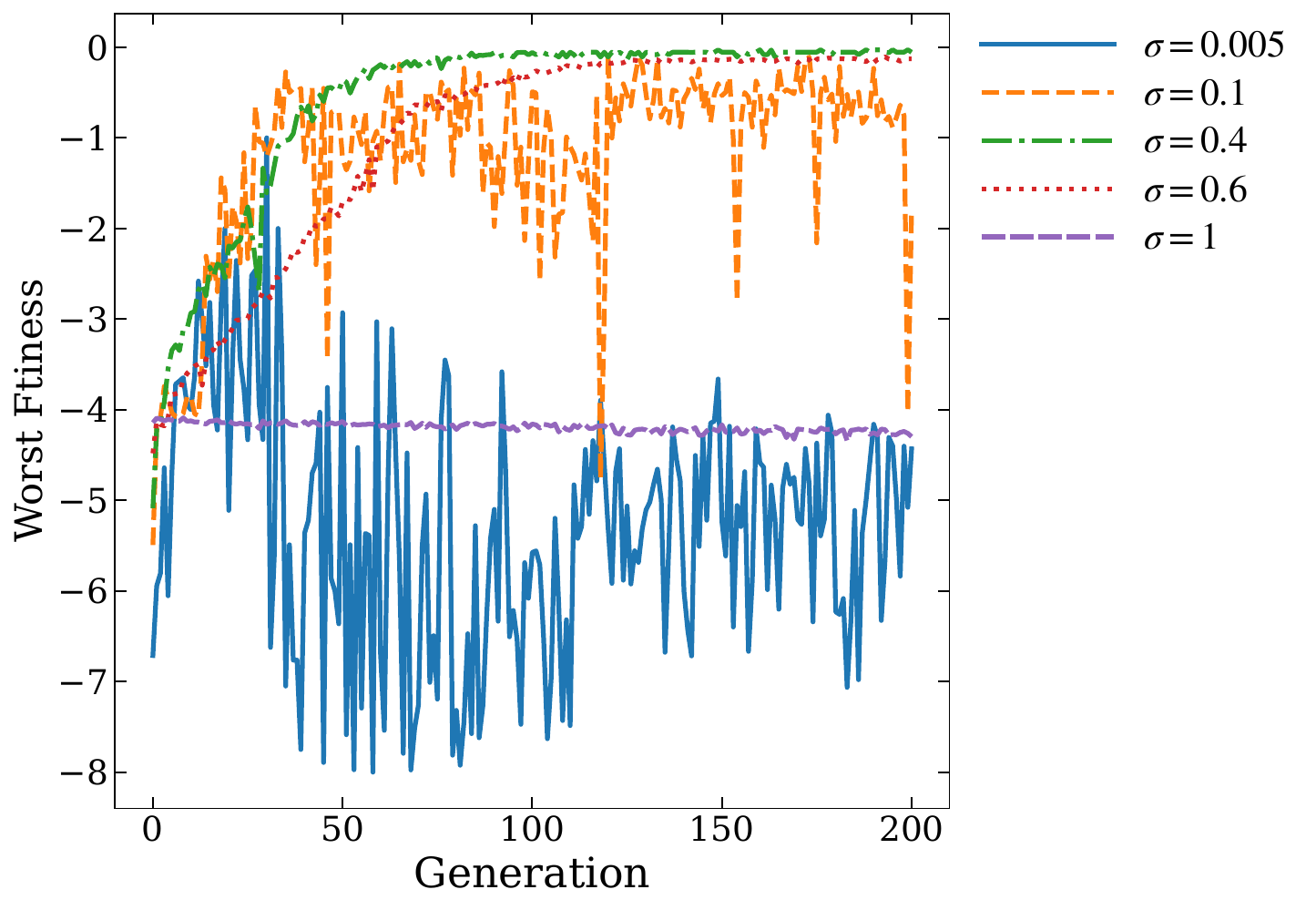}
    \small (b) Worst Fitness
  \end{minipage}
  \caption{
 Dependence on the noise strength $\sigma$ of (a) the maximum and (b) the minimum value of the mean fitness $\overline{F}(J)$ (top and worst fitness, respectively) in a population of $N(=300)$ GRNs at each generation.
 }
  \label{fig:fitness_over_generations}
\end{figure}

In Fig.~\ref{fig:fitness_over_generations} (a),
we show the top fitness in the population over generations under various values of the noise strength $\sigma$.
As the noise strength decreases, the increase in the top fitness becomes more rapid. In particular, under low-noise conditions, the maximum fitness quickly reaches its upper bound of zero. In contrast, as $\sigma$ increases, the growth of the top fitness becomes slower, and at $\sigma = 1$ it increases only weakly over generations.
The overall trend is consistent with Fig. 1(a) of the original study \cite{Kaneko2007}, reproducing \textbf{Result 1}.

\begin{figure}[ht]
  \centering
  \includegraphics[width=0.6\linewidth]{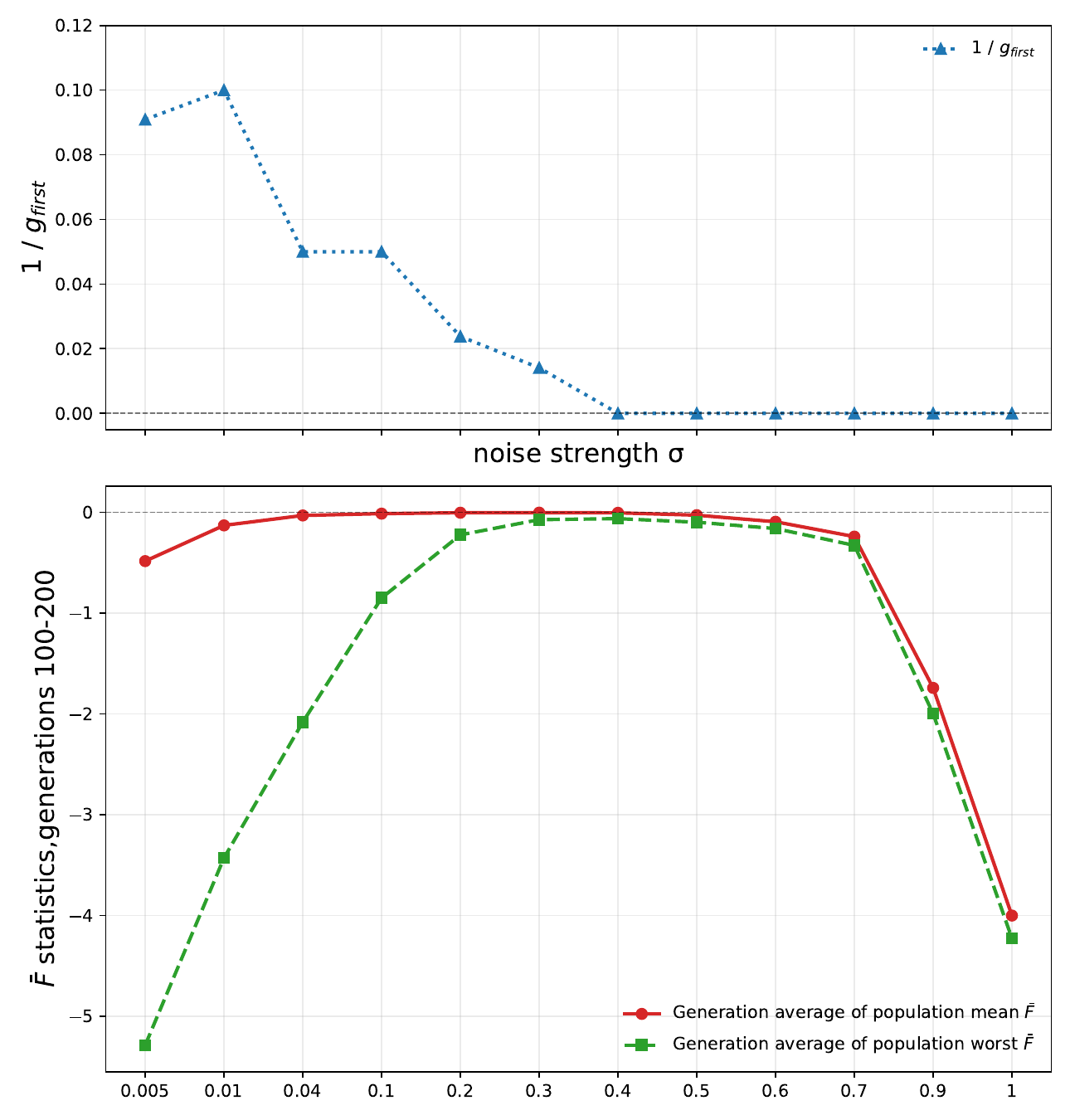}
\caption{
Evolutionary speed and population-level fitness statistics as functions of the noise strength $\sigma$.
In the upper panel, the curve with $\blacktriangle$ shows the evolutionary speed, defined as $1/g_{\mathrm{first}}$.
In the lower panel, the curves marked by $\bullet$ and $\blacksquare$ show the population-average of the mean fitness and worst fitness, respectively, as functions of noise strength $\sigma$.
Both quantities are averaged over generations 100--200.
}
  \label{fig:_fitness}
\end{figure}

To quantify the noise dependence of the number of generations required for the top fitness to reach 0, we define
the evolutionary speed  
as \(1/g_{\mathrm{first}}\), where \(g_{\mathrm{first}}\) is the first generation at which the top fitness reaches \(0\).
If the top fitness does not reach \(0\) within 200 generations, the evolutionary speed is set to \(0\).
The computed evolutionary speed is shown in the upper panel of Fig. \ref{fig:_fitness}.
The evolutionary speed is higher under low-noise conditions and decrease as the noise strength increases.
Under large-noise conditions, the top fitness do not reach \(0\) within 200 generations, resulting in an evolutionary speed of \(0\). This result is consistent with Fig.2 of the original study.

We next examine the evolution of the worst fitness in the population as shown in Fig.~\ref{fig:fitness_over_generations}(b).
Unlike the top fitness, the worst fitness remains low far from the maximum value under low-noise conditions, indicating that low-fitness individuals remain in the population throughout evolution.
As the noise strength increases, the worst fitness improves and approaches the maximum fitness value. However, when the noise becomes too large, this tendency breaks down and the overall fitness level deteriorates.
This behavior is consistent with Fig. 1(b) of the original study \cite{Kaneko2007}, reproducing \textbf{Result 2}.

Together, these results show that low-noise conditions allow rapid emergence of high-fitness individuals but tend to leave low-fitness individuals in the population.
In contrast, higher-noise conditions 
slow down the emergence of the maximum-fitness state
but facilitate improvement in the worst fitness.

The closeness between the worst and the top fitness at sufficiently large noise strengths suggests that the population-level fitness distribution should also be examined.
First, we compute population-averaged
fitness $\langle \overline{F}(J) \rangle_J$ and compare it with the worst fitness
to understand the population-level behavior.
The population-average is
calculated by averaging over individuals at each generation. These population-averaged fitness values are then averaged over generations 100--200.
As shown in bottom panel of Fig.~\ref{fig:_fitness},
the population-averaged fitness 
remains relatively high over a broad range of noise strengths up to $\sigma=0.7$.
In contrast, the worst fitness is significantly lower at small noise strengths and tends to increase as the noise strength increases up to $\sigma=0.7$.
At very large noise strengths (\(\sigma = 0.9\) and \(1.0\)),
lower fitness values are observed, suggesting that strong stochastic perturbations suppress fitness improvement.
This $\sigma$-dependence of the population-averaged fitness is consistent with Fig.2 of the original study.

\begin{figure}[ht]
  \centering
  \includegraphics[width=0.8\linewidth]{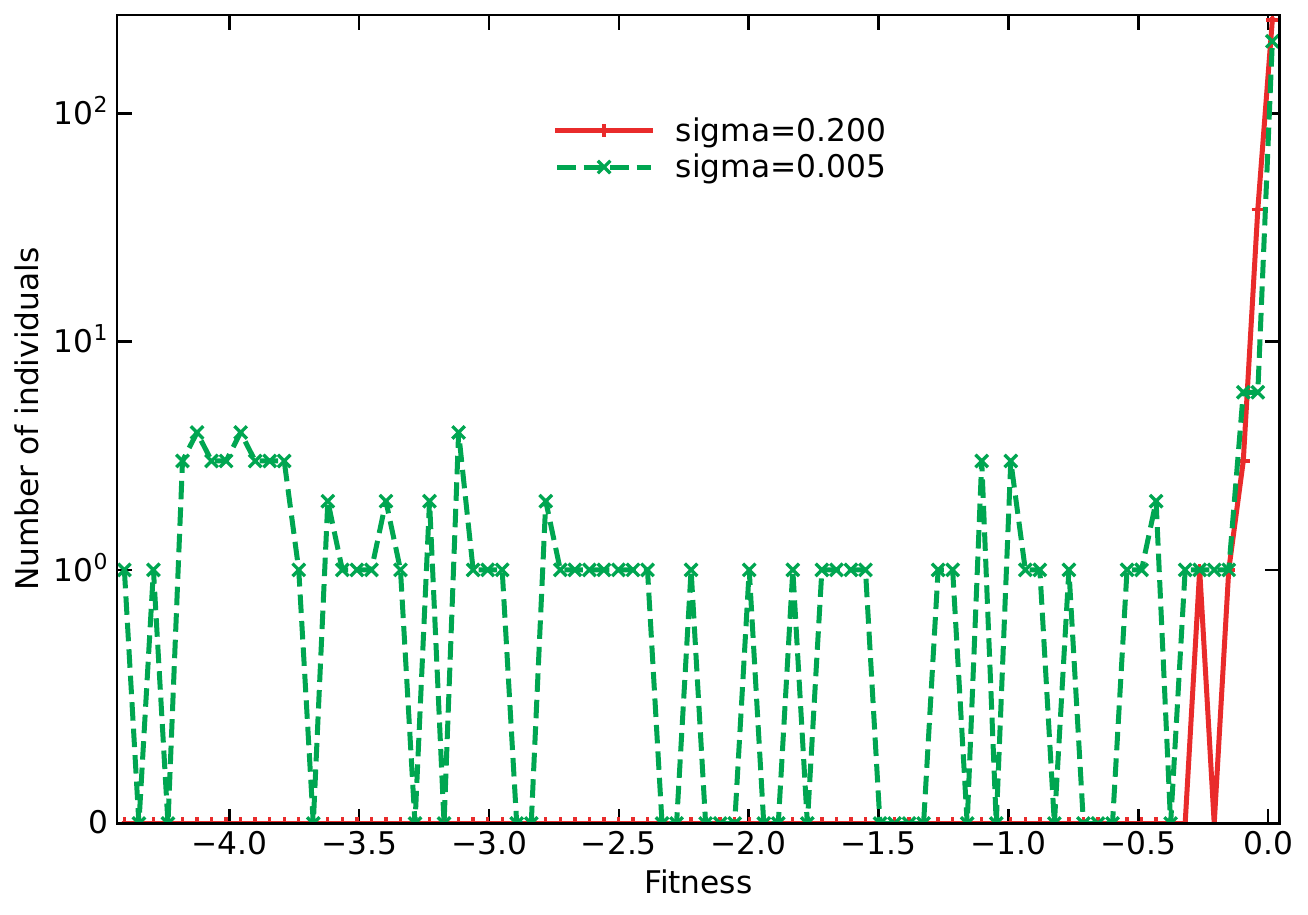}
  \caption{
Histogram of mean fitness values among individuals in the population at generation 200 under different evolutionary noise strengths.
Representative low- and intermediate-noise conditions, $\sigma=0.005$ and $\sigma=0.2$, are shown.}
  \label{fig:last_fitness_dist}
\end{figure}

For more detailed understanding of the population behavior, we next examine the histogram of fitness values in the population at the final generation under low- and sufficiently large-noise conditions.
As shown in Fig.~\ref{fig:last_fitness_dist}, 
individuals under low-noise conditions exhibit a broad range of fitness values, indicating that low-fitness individuals remain in the population, as shown in Fig.~\ref{fig:fitness_over_generations} (b), whereas under 
sufficiently large noise
conditions most individuals are concentrated near the maximum fitness value.
This behavior is consistent with Fig.3 of the original study \cite{Kaneko2007}, reproducing \textbf{Result 3}.

In addition to differences in extreme fitness values and evolutionary speed, we find clear population-level differences in the fitness distribution. 
Under low-noise conditions,
high-fitness individuals emerge rapidly, but low-fitness individuals persist, resulting in a broad fitness distribution at the final generation.
In contrast, under intermediate noise conditions, evolution proceeded more slowly, but the worst fitness improved more readily and the population becomes concentrated near higher fitness values.
These changes begin to appear around $\sigma = 0.1$--$0.2$, 
suggesting a threshold-like noise regime separating broad and concentrated population states.
This threshold-like separation qualitatively reproduces \textbf{Result~4}, although the threshold value differs from that in the original study.

\subsection{Relationship between \(V_{ip}\) and \(V_g\) }

\begin{figure}[ht]
  \centering
  \includegraphics[width=0.6\linewidth]{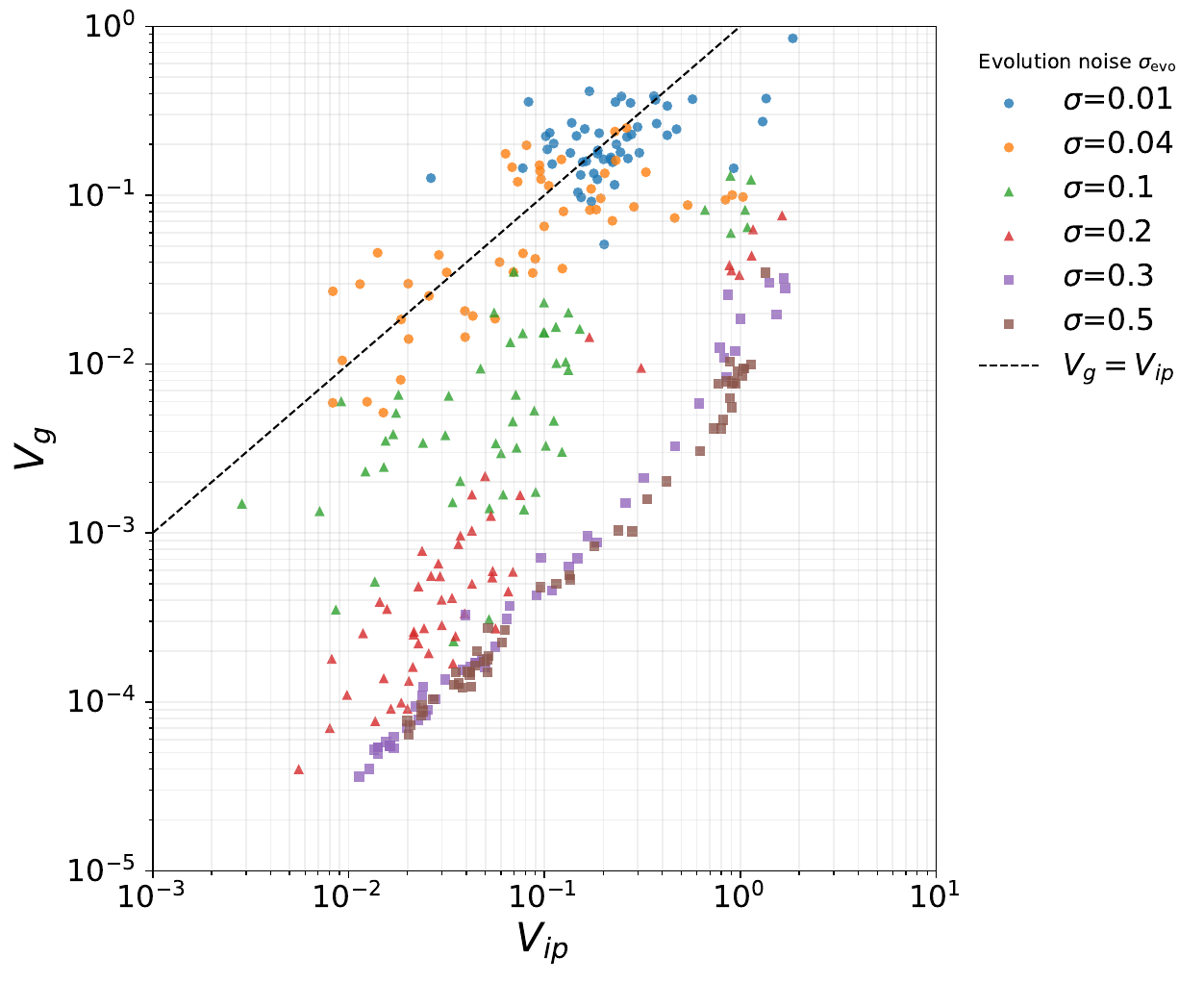}
  \caption{Evolutionary change of the relationship between \(V_g\) and \(V_{ip}\) under different noise strengths \(\sigma\).
Each point corresponds to one generation sampled every four generations. 
Circles, triangles, and squares indicate low-, intermediate-, and high-noise conditions, respectively.
As evolution proceeds, the points generally shift toward lower values of both
variances, although this tendency is weak at $\sigma=0.01$.
The dashed line indicates \(V_g=V_{ip}\).}
  \label{fig:vip_vg}
\end{figure}

We next examine the relationship between \(V_{ip}\) and \(V_g\) across noise conditions.
Fig.~\ref{fig:vip_vg}  
shows the evolutionary change of the relationship between $V_{ip}$ and $V_g$ under various noise strengths.
Data points are plotted every four generations between generations 100 and 200, and the reference line $V_{ip}=V_g$ is shown for comparison.
Under low-noise conditions, the behavior depends on the noise strength.
At the lowest noise strength, $\sigma=0.01$, both $V_{ip}$ and $V_g$
fluctuate around nearly constant values, showing little systematic decrease
during evolution.
At $\sigma=0.04$, although the noise strength is still low, both variances
decreased over generations while remaining close to the line $V_{ip}=V_g$.
Thus, in the low-noise regime, the inequality $V_{ip}>V_g$ is not clearly
established. At sufficiently large noise strengths, the decrease in the two variances becomes more
pronounced.
In particular, $V_g$ is more strongly suppressed than $V_{ip}$, and the
number of generations satisfying $V_{ip}>V_g$ generally increases, particularly
for $\sigma \gtrsim 0.1$.
Furthermore, at even higher noise strengths,
a clearer proportional relationship between $V_{ip}$ and $V_g$ emerges.
Our observation reproduces \textbf{Result 5}, and is qualitatively consistent with Fig.4 of the original study.

\subsection{Robustness of evolved GRNs}

We next compared the properties of gene-expression dynamics defined on evolved GRNs.
For each evolutionary noise condition, one representative network, namely the individual with the highest mean fitness, was chosen from the population at the final generation.
When multiple individuals achieved \(\overline{F}(J)=0\) under low-noise conditions, one of them is selected at random.
When no individual reached \(\overline{F}(J)=0\) by generation 200 under high-noise conditions, the individual with the highest mean fitness is selected.

\subsubsection{Robustness to noise in expression dynamics}
\label{sec:robustness_to_noise}

To characterize the dynamics defined by the evolved GRNs, we numerically simulate the gene-expression dynamics \eqref{eq:grn_dynamics} on the representative GRNs.
The simulations were carried out under various noise levels $\sigma_{\mathrm{sim}}$, which is not necessarily equal to the noise level $\sigma_{\mathrm{evo}}$ under which the GRNs evolved.
Varying $\sigma_{\mathrm{sim}}$ enables us to test whether an evolved GRN can still achieve the high-fitness expression under noise levels different from the evolutionary condition. 
This provides a measure of the robustness of the evolved gene-expression dynamics to changes in the level of stochastic fluctuations.

The numerical settings for the simulations of the gene-expression dynamics on the evolved representative GRNs are as follows.
For each representative network, 10,000 trials of the expression dynamics \eqref{eq:grn_dynamics} are performed from the fixed initial condition \(x_i(0)=-1\) for all \(i\).
This fixed initial condition is used to isolate the effect of stochastic noise during the dynamics.
The fitness is calculated using the same $T_{\mathrm{ini}}$ and $T_{\mathrm{f}}$.
As a measure of robustness, we compute the fraction of trials that reach ${F}(J)=0$, which we refer to as success rate.

\begin{figure}[ht]
  \centering
  \includegraphics[width=0.8\linewidth]{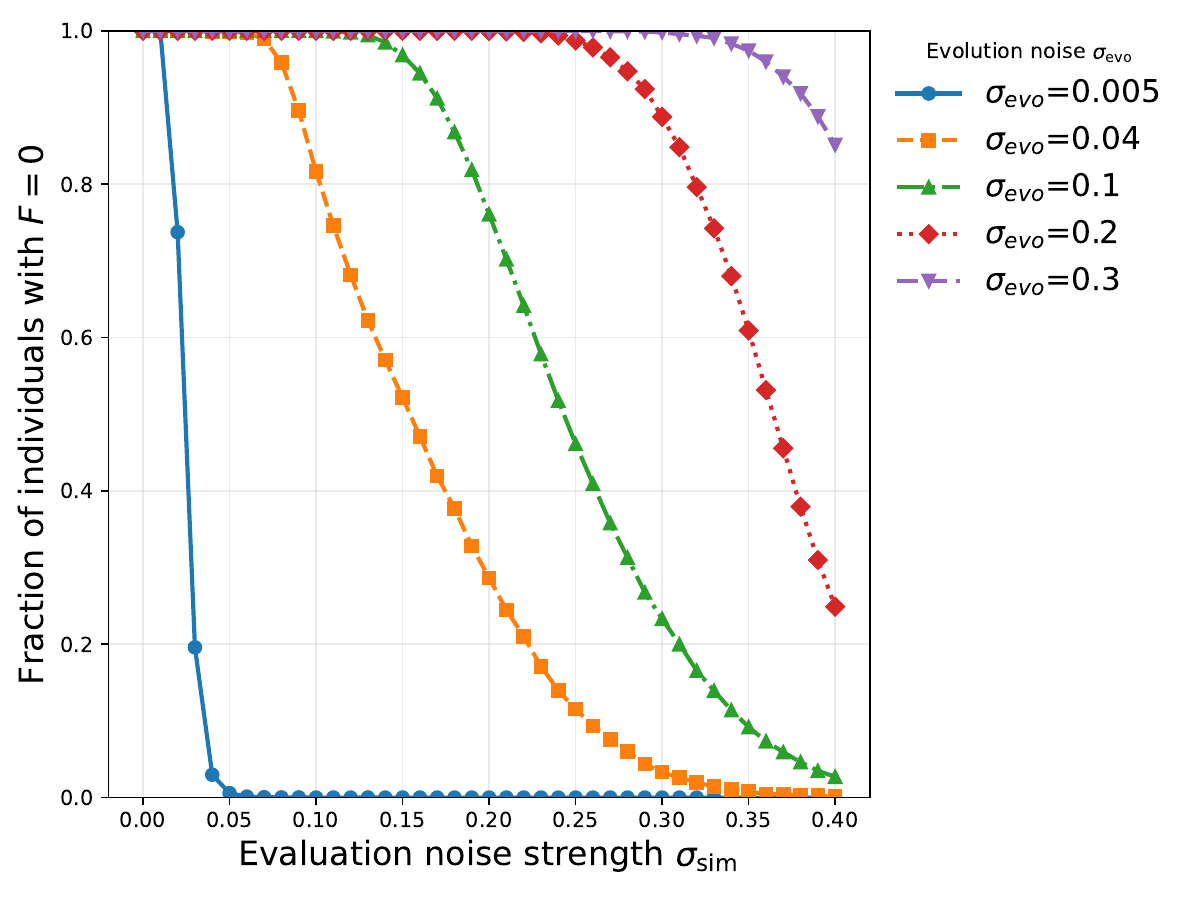}
  \caption{
Fraction of trials reaching $F=0$ (success rate) as a function of evaluation noise strength \(\sigma_{\mathrm{sim}}\).
For each representative individual and each value of \(\sigma_{\mathrm{sim}}\), the expression dynamics were simulated 10,000 times.}
  \label{fig:under_noise_fitness}
\end{figure}

In Fig.~\ref{fig:under_noise_fitness}, the success rates are shown as a function of $\sigma_{\mathrm{sim}}$ for various $\sigma_{\mathrm{evo}}$.
As shown, the success rates decrease as $\sigma_{\mathrm{sim}}$ increases, but the manner of decrease depends on $\sigma_{\mathrm{evo}}$.
For small $\sigma_{\mathrm{evo}}$, e.g., $\sigma_{\mathrm{evo}} = 0.005$, the success rate drops sharply once $\sigma_{\mathrm{sim}}$ exceeds the noise level experienced during evolution, $\sigma_{\mathrm{evo}}$. 
In contrast, for larger $\sigma_{\mathrm{evo}}$, the networks maintain $F(J)=0$ not only for $\sigma_{\mathrm{sim}} \lesssim \sigma_{\mathrm{evo}}$, but also for substantially larger values of $\sigma_{\mathrm{sim}}$.

In other words, a success rate of unity indicates that GRNs evolved under a given noise strength can reliably reach the target gene-expression state. Conversely, networks evolved under sufficiently large noise levels exhibit the highest tolerance to increases in evaluation noise, maintaining high fitness even when $\sigma_{\mathrm{sim}} > \sigma_{\mathrm{evo}}$. This result is qualitatively consistent with Figure 5 of the original study.

\subsubsection{Robustness to initial condition}
\label{sec:basin_expansion}

The expression dynamics defined on GRNs may also depend on the choice of initial conditions.
In the evolutionary simulations, initial expression states were initialized at $x_i(0)=-1$ for all $i$.
Accordingly, it is possible that trajectories from this initial condition toward the target  
states are shaped through evolution.
Given GRNs evolved under this setting, it is natural to ask how their dynamical behavior changes when the initial conditions are varied.
Such dependence reflects the structure of the state space, including the presence and organization of multiple attractors. If a GRN can reach a target state from a wide range of initial conditions, it can be regarded as robust with respect to initial condition dependence.

The settings for examining the initial condition dependence are as follows.
For each representative network, 
selected in the same manner as in Sec.~\ref{sec:robustness_to_noise} under each evolutionary noise condition,
we performed 10,000 trials with randomly sampled initial expression states. 
In this analysis, the stochastic noise in the gene-expression dynamics was set to zero in order to isolate the effect of initial conditions. 
Accordingly, the deterministic gene-expression dynamics were integrated using the fourth-order Runge-Kutta method with \(dt=0.05\).
We quantify the dependence on initial conditions using the distribution of fitness values. 
If this distribution is sharply peaked  
near the target fitness values
with a small variance, the system is considered robust to initial conditions. 
In contrast, if the distribution is broad or peaked at low fitness, the system exhibits strong dependence on initial conditions and is not robust.

\begin{figure}[ht]
  \centering
  \includegraphics[width=0.8\linewidth]{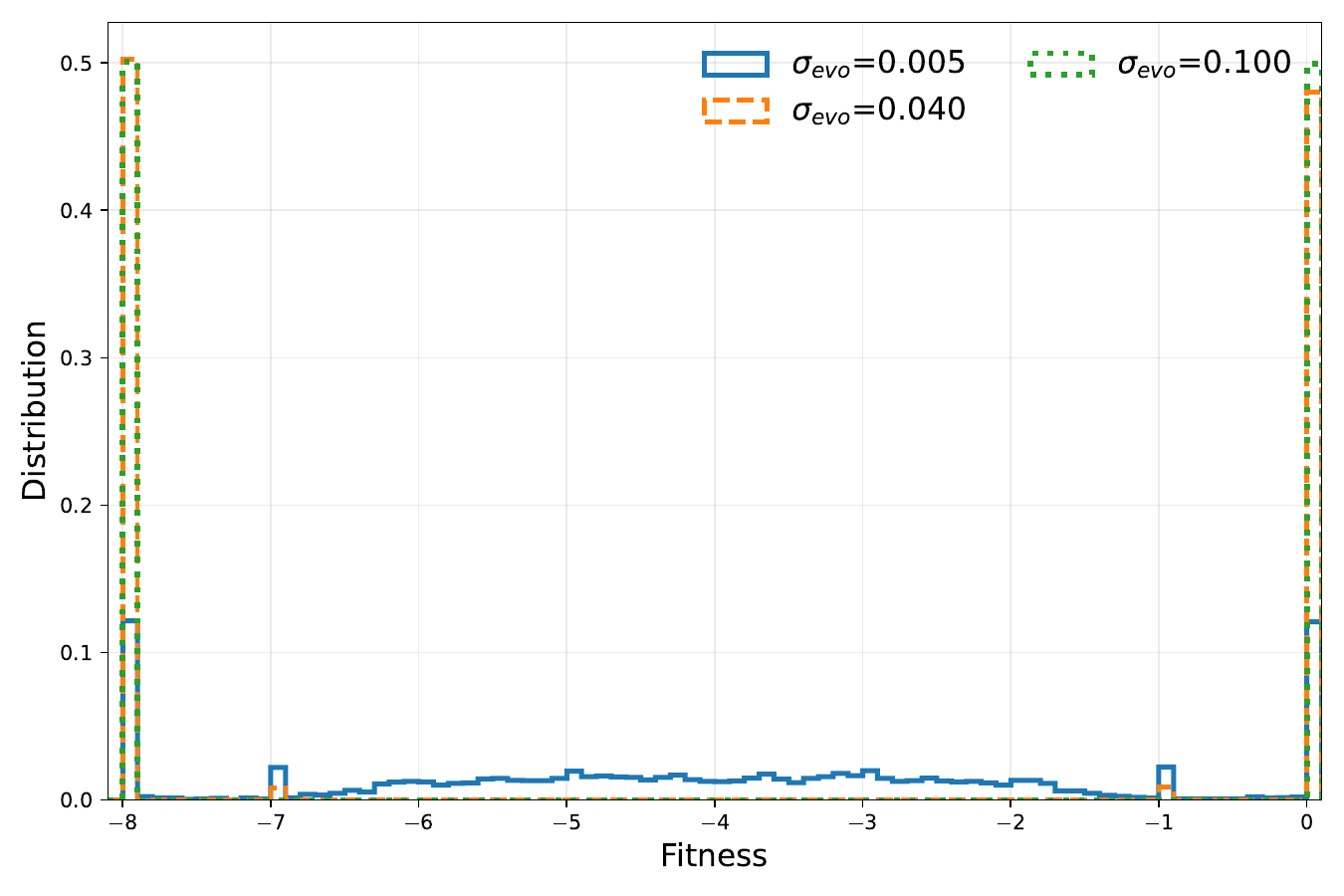}
\caption{
Fitness distributions obtained from random initial conditions for representative networks evolved under different noise strengths.
The bin width is set to 0.1.
}
  \label{fig:fitness_distribution}
\end{figure}

Fig.~\ref{fig:fitness_distribution} shows the histogram of the fitness values for three representative values of expression noise $\sigma_{\mathrm{evo}}$.
Here, the target state is defined as the highest-fitness state with \(F=0\).
Due to the symmetry of the dynamics, the state corresponding to \(F=-8\) is equivalent to this target state and is also regarded as a target state.
Accordingly, states other than $F=0$ or $F=-8$ are regarded as non-adapted.
As shown in Fig.\ref{fig:fitness_distribution}, for $\sigma = 0.005$, the fitness distribution is broad, indicating that many trials do not reach the target phenotype; only about 24\% of trials achieve the target states. 
For $\sigma = 0.04$, small secondary peaks appear at non-target fitness values, although most trials still reach the target states. 
Finally, for $\sigma = 0.1$, the distribution becomes sharply peaked at $F=0$ and $F=-8$, indicating that almost all trajectories converge to the target states.

These results suggest that the  basin structure depends on the evolutionary noise level.
For the network evolved under very low noise, the final phenotype strongly depends on the initial condition.
In contrast, networks evolved under larger noise strengths reach the target states from a larger fraction of random initial conditions.
Taken together, the results of Secs.~\ref{sec:robustness_to_noise} and \ref{sec:basin_expansion} qualitatively reproduce Result 6: evolution under sufficiently large noise leads both to enhanced robustness of gene-expression dynamics and to an expansion of the basin of attraction of the target states 
\footnote{In the original study, the emergence of the relationship \(V_{ip}>V_g\) and the expansion of the basin of attraction were observed at a similar noise strength. In our simulations, the correspondence between these two properties is not as clear as that reported in the original study, which may be attributed to differences in simulation conditions.}

\section{Additional Simulation Results: Robustness of Evolved Populations to Initial Conditions}

The representative-network analysis in the previous section suggests basin expansion under sufficiently large evolutionary noise.
However, because only one representative network is analyzed for each evolutionary noise condition, it remains unclear whether this tendency is typical among evolved high-fitness individuals.
To address this point, we next perform a population-level analysis of robustness to initial conditions.

We select high-fitness individuals from the evolved population for each noise condition. For \(\sigma \le 0.20\), individuals with evolutionary fitness \(F=0\) were extracted. For \(\sigma \ge 0.30\), few individuals reach \(F=0\) during evolution because of the high noise level. 
However, because the population consists almost entirely of individuals with sufficiently high fitness, all 300 individuals are regarded as sufficiently evolved and are included in the analysis.

For each 
selected individual, 10,000 trials of the expression dynamics are performed from randomly chosen initial expression states. The numerical conditions for the simulation are the same as those in Sec.\ref{sec:basin_expansion}. The diversity of the resulting expression states is quantified using two indices: the {\it invalid count} \(IC\) and the {\it unique fitness count} \(UF\). The invalid count \(IC\) is defined as the number of trials whose resulting fitness is neither \(F=0\) nor \(F=-8\).
Hence, individuals with $IC=0$ are completely robust to initial-state variation.
The unique fitness count \(UF\) is defined as the number of distinct fitness values, after binning 
with a bin width of 0.1.
Because of
the symmetry of this model, individuals with \(IC=0\) necessarily have \(UF=2\). Therefore, 
\(UF\) is evaluated only for individuals with \(IC>0\). The number of individuals with $IC=0$ is recorded as \(N_{IC=0}\).

Table~\ref{tab:icuf_high} summarizes the number of analyzed individuals \(N_{\mathrm{sel}}\), the number of individuals with \(IC=0\) (\(N_{IC=0}\)), 
and summary statistics of \(IC\) and \(UF\): the median (Med.), third quartile (Q3) and maximum value (Max), for individuals with \(IC>0\) under each noise condition.
Overall, the number of individuals with \(IC=0\) tends to increase as the evolutionary noise strength \(\sigma\) increases.
For individuals with \(IC>0\), the median, third quartile, and maximum of \(IC\) also generally decrease as $\sigma$ increases, 
although a non-monotonic behavior is observed at $\sigma=0.5$.
This overall trend indicates that the distribution of IC shifts toward lower values with increasing $\sigma$, suggesting that the target state can be reached from a broader range of initial conditions, even among individuals with $IC>0$.
Similarly, the summary statistics
of \(UF\) decrease overall, indicating that the diversity of final fitness values reached from different initial states becomes smaller, 
even among individuals that do not always achieve $F=0$ or $F=-8$ from all initial conditions.
In contrast to the previous section, which focused only on individuals with the highest fitness, the present results show that similar trends are observed across the broader population, suggesting population-level robustness.

These trends were not strictly monotonic with respect to the noise strength.
Nevertheless, they overall indicate
that individuals evolved under higher expression noise conditions are less likely to reach inappropriate final states from random initial conditions and tend to 
exhibit simpler fitness distributions across initial states.
Taken together with the representative-network analysis in the previous section, these population-level results further support the basin-expansion interpretation of Result 6.

\begin{table}[t]
\caption{Number of selected high-fitness individuals ($N_{\mathrm{sel}}$), number of individuals with $IC=0$ ($N_{IC=0}$), and summary statistics of $IC$ and $UF$ among individuals with $IC>0$.}
\label{tab:icuf_high}
\begin{ruledtabular}
\begin{tabular}{c c c c c c c c c}
& & & \multicolumn{3}{c}{$IC$} & \multicolumn{3}{c}{$UF$} \\
\cline{4-6}\cline{7-9}
$\sigma$ & $N_{\mathrm{sel}}$ & $N_{IC=0}$ & Med. & Q3 & Max & Med. & Q3 & Max \\
0.005 & 207 & 0 & 8080 & 8680 & 9343 & 81 & 81 & 81 \\
0.01  & 203 & 0 & 2025 & 2808 & 5248 & 72 & 79 & 81 \\
0.04  & 245 & 0 & 867  & 1345 & 3008 & 11 & 15 & 55 \\
0.10  & 238 & 6 & 118  & 337  & 1198 & 7  & 9  & 30 \\
0.20  & 203 & 4 & 137  & 257  & 3813 & 7  & 15 & 41 \\
0.30  & 300  & 75 & 45   & 79   & 537  & 4  & 5  & 14 \\
0.40  & 300  & 34 & 8    & 18   & 670  & 5  & 6  & 29 \\
0.50  & 300   & 2 & 117  & 207  & 868  & 8  & 9  & 17 \\
0.60  & 300  & 50 & 5    & 18   & 70   & 5  & 6  & 11 \\
0.70  & 300 & 153 & 2    & 4    & 18   & 4  & 4  & 8 \\
\end{tabular}
\end{ruledtabular}
\end{table}

\section{Conclusion}

In this study, we have reimplemented the gene regulatory network evolution model described by Kaneko~\cite{Kaneko2007} and examined how phenotypic noise affects the evolution of robustness.
The results have shown that phenotypic noise shapes not only the speed of evolution but also the distribution of fitness in the evolved population.
Specifically, low-noise conditions allowed low-fitness individuals to persist, 
whereas sufficiently large but not excessive noise conditions led the population to concentrate near high fitness.
This noise-dependent trend is consistent with the original study.

The relationship between \(V_{ip}\) and \(V_g\) also showed a trend consistent with the original study.
Under low-noise conditions, the relation \(V_{ip}>V_g\) was not clearly observed, whereas above a certain noise strength, \(V_{ip}>V_g\) tended to hold throughout the evolutionary process.
The noise region in which this relation became clear roughly corresponded to the boundary region where low-fitness individuals became less likely to remain in the population.
Thus, the present results support the interpretation of the original study: sufficiently large phenotypic noise
suppresses the accumulation of the networks with unstable expression dynamics during selection and promotes the evolution of networks robust to both noise and mutation.

Although the qualitative \(V_{ip}\)--\(V_g\) relationship was reproduced, the boundary noise strength appeared at a higher value in the present study than in the original study.
One possible explanation is that differences in simulation settings, including the mutation scheme, may affect the evolutionary dynamics and shift the noise strength at which \(V_{ip}>V_g\) emerges.
Because the present model describes a nonequilibrium evolutionary process, the population structure generated by mutation-selection dynamics may depend on details of the evolutionary rules.
Such differences may contribute to the observed shift in the balance between \(V_g\) and \(V_{ip}\).
Such differences may contribute to the shift in the noise strength at which the relationship $V_{ip}>V_g$ emerges.
Further investigation is needed to determine how simulation settings influence the emergence of the $V_{ip}-V_g$ relationship.

In addition to the analysis of representative networks, we performed a random-initial-condition analysis for the evolved population.
The results showed that, even among individuals with comparable fitness during evolution, those evolved under larger noise strengths were more likely to reach a target state starting from random initial conditions.
Their final-fitness distributions were also narrower and unimodal, apart from a trivial splitting into two symmetric peaks caused by a symmetry inherent to the model.
These findings indicate that the change in basin structure discussed in the original study is not limited to a small number of representative individuals, 
but is observed across a broad range of evolved individuals.

In this study, we focused on robustness toward a single target phenotype.
However, biological systems often need to switch between multiple phenotypes in response to environmental conditions or external inputs.
Recent evolutionary models have shown that low-dimensional switching paths between phenotypes can emerge through evolution~\cite{SakataKaneko2023}, and that evolutionary robustness can constrain high-dimensional phenotypic changes into a few dominant degrees of freedom~\cite{FurusawaKaneko2018}.
Together with the present results, these previous studies suggest that the present model could be extended to study the relationship among robustness, plasticity, and low-dimensional structure in phenotype space.
In future work, we aim to analyze models with multiple target phenotypes or changing target phenotypes, and to investigate how robustness and plasticity interact during evolution.

\begin{acknowledgments}
The authors would like to thank Tetsuo Deguchi, Yuki Izumida and Kunihiko Kaneko for helpful discussions and comments.
\end{acknowledgments}

\bibliographystyle{apsrev4-2}
\bibliography{references}

\end{document}